\documentclass{ptapap}
\usepackage[]{natbib}

\author{Konstanze Zwintz}[konstanze.zwintz@uibk.ac.at,UIBK]
\affil[UIBK]{Universit\"at Innsbruck, Institute for Astro- and Particle Physics\\
  Technikerstrasse 25/8, A-6020 Innsbruck, Austria}

\title{A BRITE view on $\delta$ Scuti and $\gamma$ Doradus stars}

\begin{document}

\maketitle

\begin{abstract}

BRITE-Constellation has obtained data for a few $\delta$ Scuti and $\gamma$ Doradus type stars. A short overview of the pulsational content found in five stars -- $\beta$ Cassiopeiae, $\epsilon$ Cephei, M Velorum, $\beta$ Pictoris and QW Puppis -- is given and the potential of BRITE-Constellation observations of $\delta$ Scuti and $\gamma$ Doradus pulsators is discussed.

\end{abstract}

\section{Introduction}

$\delta$ Scuti stars have spectral types A to early F, masses between $\sim$ 1.5 and 4.0 solar masses and pulsation periods between 18 minutes and 6 hours. They typically show multiple radial and non-radial pressure ($p$) mode frequencies of low degree $\ell$\,\, which are excited by the $\kappa$-mechanism acting in the second Helium ionization zone \citep[e.g.,][]{aerts10}. Although $\delta$ Scuti stars are one of the first groups of stars discovered to pulsate, they are still one of the least understood. Open questions include for example observed changes in pulsation amplitudes \citep[e.g.,][]{bowman16}, hybrid pulsators showing gravity ($g$) and pressure ($p$) modes \citep[e.g.,][]{kurtz14}, spots on the stellar surface, the presence of magnetic fields \citep[e.g.,][]{neiner15,escorza16}, differential rotation of core versus envelope \citep[e.g.,][]{kurtz14}, or the influence of the sometimes rapid rotation on the pulsation frequencies \citep[e.g., ][]{reese09}. In only few cases, $\delta$ Scuti stars show characteristic spacings and splittings that can be used for an asteroseismic interpretation \citep[e.g.,][]{zwintz11,zwintz13}.

$\delta$ Scuti pulsation has also been observed in several pre-main sequence (pre-MS) objects \citep[e.g.,][]{zwintz14} that gain their energy mainly from gravitational contraction on their way from the birthline to the zero-age main sequence (ZAMS). For pre-MS $\delta$ Scuti stars a relation between their pulsational properties and the relative stage in their pre-MS evolution could be discovered: The least evolved stars that are still close to the birthline and, hence, have the largest radii, show the slowest pulsations. The further the pre-MS pulsators have progressed in their evolution to the ZAMS, and, hence, the more compact they have become, the faster they oscillate \citep{zwintz14}. 


$\gamma$ Doradus type stars have spectral types from late A to early F and show $g$-mode oscillations driven by the convective flux blocking mechanism \citep{guzik00}.
Their pulsation periods range from $\sim$0.3 to 3 days \citep[e.g.,][]{kaye99} where modes with same spherical degree $\ell$ and different radial orders $n$ show equidistant spacings in period \citep{tassoul80}. If a chemical gradient is present at the edge of the convective core, the $g$-mode resonance cavity is modified, hence, resulting in periodic dips in the period spacing pattern \citep{miglio08}. Using four-year high-precision space-based photometric time series obtained by the Kepler \citep{koch10} satellite, recently period spacing patterns could be detected \citep{vanreeth15aa,vanreeth15apjs} and subsequently used to determine the rotation rates and to identify mainly prograde dipole gravity and gravito-inertial modes from the pulsation frequencies \citep{vanreeth16}.

With masses between $\sim$1.4 to 2.5 solar masses, $\gamma$ Doradus stars are the perfect test-beds to study the transition range from low-mass stars with radiative cores and convective envelopes to high-mass stars with convective cores and radiative envelopes. Hence, asteroseismology of $\gamma$ Doradus pulsators allows to calibrate and improve upon our existing theories of stellar structure and evolution in the mass range of $\sim$1 to 2 solar masses. 


\section{BRITE observations of $\delta$ Scuti and $\gamma$ Doradus stars}

The main observational focus of BRITE-Constellation lies on the most massive O and B type stars. Only $\sim$6\% of the targets have spectral types A and only $\sim$12\% are of spectral types F, and not all of them have BRITE-Constellation data of sufficient quality for a pulsational analysis, e.g., because the data set length is too short for a thorough analysis.

As of August 2016, BRITE-Constellation has obtained data for seven $\delta$ Scuti stars and two $\gamma$ Doradus stars for which I am the contact-PI. The data sets for the two $\delta$ Scuti pulsators 16 Persei and 44 Tauri have been insufficient for a pulsational analysis and, hence, had to be omitted. The analysis of the $\gamma$ Doradus star 43 Cygni is described separately by S. G\"ossl in these proceedings. Here I will describe the latest findings of four stars classified as $\delta$ Scuti type (i.e., $\beta$ Cassiopeiae, $\epsilon$ Cephei, M Velorum, $\beta$ Pictoris) and one as $\gamma$ Doradus pulsator (i.e., QW Puppis) using BRITE-Constellation data (see Table \ref{obs}).

Data reduction has been performed using the tool developed by M. Kondrak (described in these proceedings) including decorrelations with CCD temperature, and the X and Y positions of the point-spread function on the CCD. 
For the frequency analyses, we used the software package {\sc Period04} \citep{period04} that combines Fourier and least-squares algorithms. Frequencies were then prewhitened and considered to be significant if their amplitudes exceeded four times the local noise level in the amplitude spectrum \citep{breger93,kuschnig97}.
We verified the analysis using the SigSpec software \citep{reegen07}. SigSpec computes significance levels for amplitude spectra of time series with arbitrary time sampling. The probability density function of a given amplitude level is solved analytically and the solution includes dependences on the frequency and phase of the signal.

\begin{table}[htb]
\caption{Overview of the BRITE-Constellation observations of the here discussed $\delta$ Scuti and $\gamma$ Doradus stars: time bases (T) in days obtained by BRITE-Austria (BAb), Lem (BLb), Uni-BRITE (UBr), Heweliusz (BHr) and / or BRITE-Toronto (BTr). Data sets of insufficient quality for a pulsational analysis are marked with an asterisk ($^{\star}$).}
\label{obs}
\begin{center}
\begin{tabular}{cccccc}
\hline
\hline
\multicolumn{1}{l}{star} & \multicolumn{1}{c}{BAb}& \multicolumn{1}{c}{BLb} & \multicolumn{1}{c}{UBr} & \multicolumn{1}{c}{BHr} & \multicolumn{1}{c}{BTr}  \\
\multicolumn{1}{c}{ } & \multicolumn{1}{c}{T [d]} & \multicolumn{1}{c}{T [d]}  & \multicolumn{1}{c}{T [d]} & \multicolumn{1}{c}{T [d]} & \multicolumn{1}{c}{T [d]} \\
\hline
$\beta$ Cassiopeiae & 58 & 6 & - & 47 &  44 \\
$\epsilon$ Cephei & - & 18$^{\star}$ & - & 24$^{\star}$ & 48 \\
M Velorum & 37$^{\star}$ & - & - & - & 52 \\
QW Puppis & - & - & - & 78 & - \\
$\beta$ Pictoris & - & - & - & 78 & - \\
\hline
\end{tabular}
\end{center}
\end{table}

\subsection{$\beta$ Cassiopeiae (HR 21, HD 432)}

The F2 star $\beta$ Cassiopeiae is reported to be a radial $\delta$ Scuti pulsator with a single pulsation period of 2.5 hours that has already evolved towards the Terminal Age Main Sequence \citep[TAMS; ][]{riboni94}.
\citet{che11} determined a rotation period of 1.12 $\pm$ 0.04 d from interferometry. Together with a projected rotational velocity, $v sin i$, of 70 $\pm$ 1 kms$^{-1}$, the authors conclude that the star rotates with 92\% of its critical velocity and must be seen close to pole-on. Hence, $\beta$ Cas' radius is $\sim$24\% greater at the equator than at the poles and its effective temperature, $T_{\rm eff}$, is $\sim$1000\,K higher on the poles than at the equator \citep{che11}.

\begin{figure}
\centering
\includegraphics[width=0.8\textwidth]{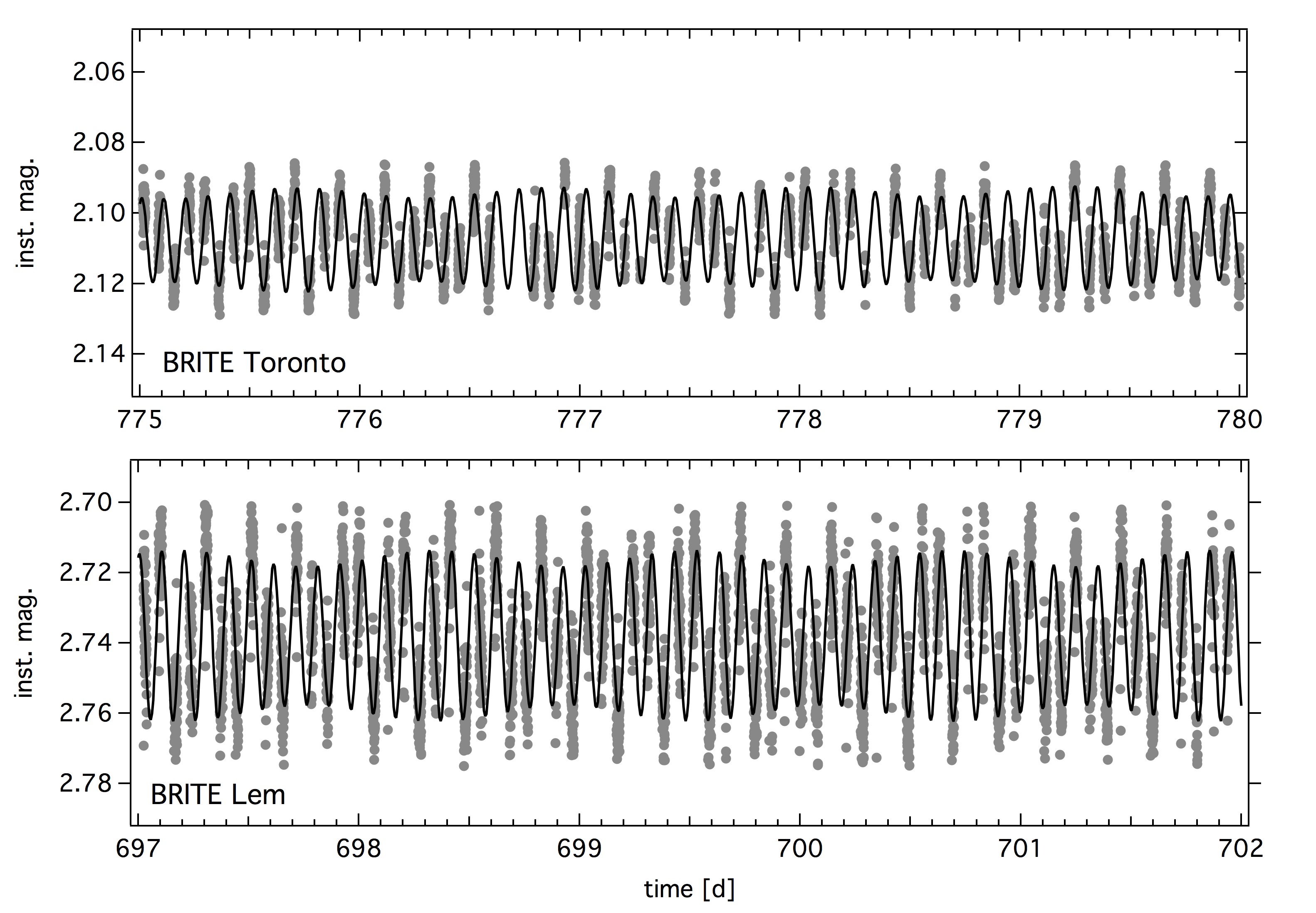}
\caption{$\beta$ Cas: 5-days subsets of the red filter BRITE-Toronto (upper panel) and the blue filter Lem (lower panel) light curves and the corresponding multi-sine fit using the two frequencies at 9.897\,d$^{-1}$ and 9.043\,d$^{-1}$ (black lines).}
\label{betacas}
\end{figure}

BRITE-Constellation obtained data with different time bases during the CasCep-I field observations in 2015 using BRITE-Austria, Lem, Heweliusz and BRITE-Toronto (see Table \ref{obs}.)
Figure \ref{betacas} shows 5-day subsets of the photometric time series obtained by BRITE-Toronto in the red filter (upper panel) and by Lem in the blue filter (lower panel).
The frequency analysis yielded two close frequencies at 9.897\,d$^{-1}$ and 9.043\,d$^{-1}$ whose difference amounts to 0.860\,d$^{-1}$ or a period of 1.16 d. This coincides well with the rotation period found from interferometry \citep[as mentioned above;][]{che11}. Further analysis is ongoing and will be subject of a future publication.

\subsection{$\epsilon$ Cephei (HR 8494, HD 211336)}

Using photometric time series obtained from space with the WIRE satellite, \citet{bruntt07} identified 26 $\delta$ Scuti pulsation frequencies for $\epsilon$ Cephei in the frequency range between 12.7 and 34\,d$^{-1}$. The observed regular spacings found in the data were attributed to successive radial orders \citep{breger09}. 

$\epsilon$ Cephei was included in the BRITE-Constellation CasCep-I field (2015) and observed with Lem, Heweliusz and BRITE-Toronto. Unfortunately, the quality of the data obtained by Lem and Heweliusz is insufficient for an analysis of the star's pulsations. BRITE-Toronto observed $\epsilon$ Cephei for $\sim$48 days. From this data set we identified 24 frequencies in the range from 5 to 30\,d$^{-1}$ and four frequencies between 0 and 1.5\,d$^{-1}$. The seismic interpretation and comparison to previous works is ongoing and will include a detailed analysis of the spectroscopic time series observations obtained from ground.

\subsection{M Velorum (HD 83446)}

No information about the variability of M Velorum can be found in the literature. 

\begin{figure}
\centering
\includegraphics[width=0.8\textwidth]{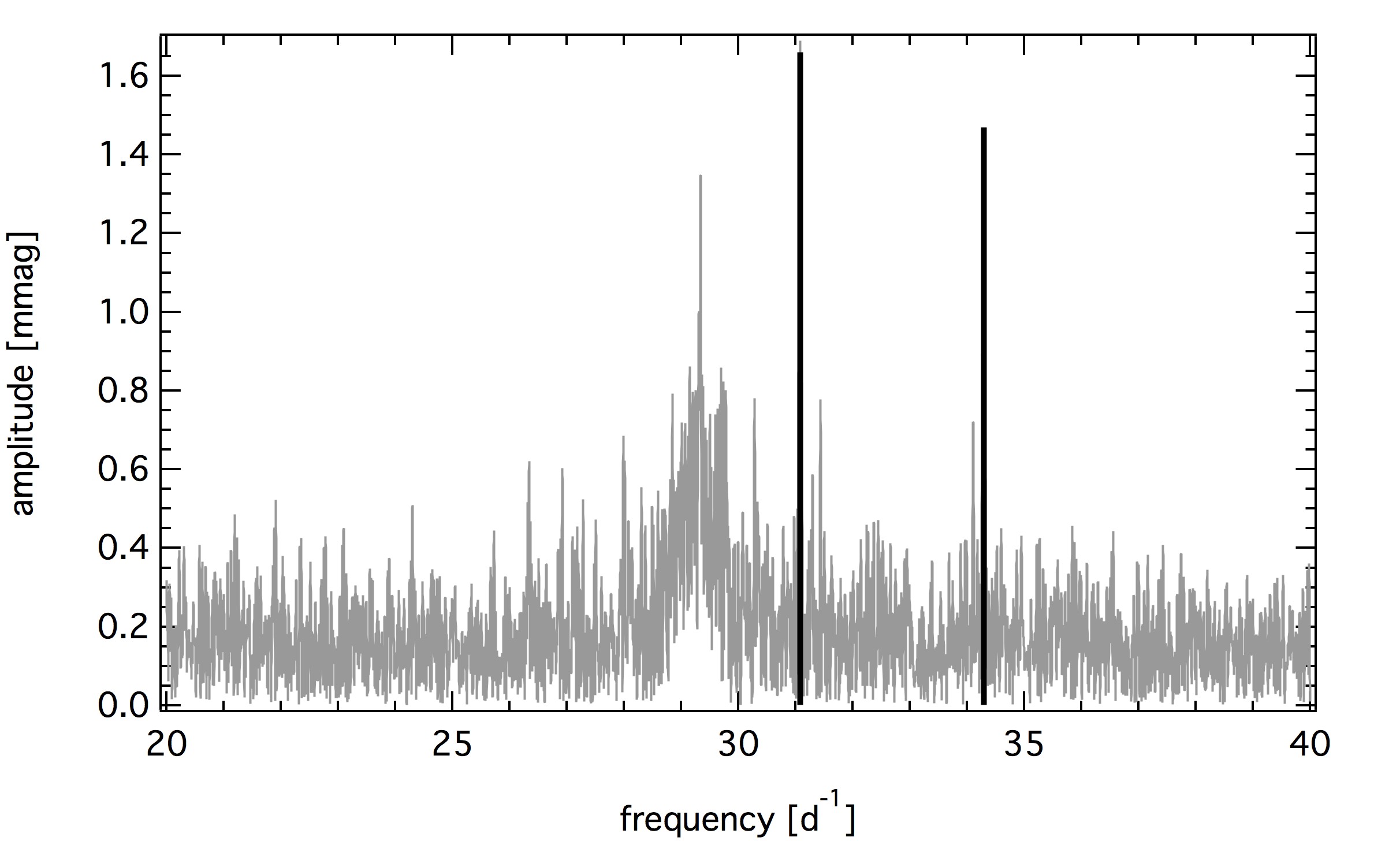}
\caption{Amplitude spectrum of the BRITE-Toronto data for M Velorum (grey), where the discovered two pulsational frequencies are marked in black.}
\label{mvel}
\end{figure}

BRITE-Toronto observed M Velorum for $\sim$ 52 days during the VelaPuppis-II field observing run from December 2014 to May 2015. The data obtained by BRITE-Austria during the same time are unfortunately insufficient for an asteroseismic analysis and had to be omitted. The BRITE-Toronto data allowed us to identify M Velorum as $\delta$ Scuti type pulsating star with two pulsation frequencies: f1 at 31.0806\,d$^{-1}$ with an amplitude of 1.6 mmag and f2 at 34.2908\,d$^{-1}$ with an amplitude of 1.5 mmag. M Velorum's amplitude spectrum is shown in Figure \ref{mvel}. Note that the peak around 29\,d$^{-1}$ corresponds to twice the orbital frequency of the satellite.

\subsection{QW Puppis (HD 55892)}

With a $V$ magnitude of 4.49 mag, QW Puppis is the third brightest known $\gamma$ Doradus type pulsator. \citet{poretti97} reported four frequencies at 1.0434\,d$^{-1}$, 0.9951\,d$^{-1}$, 1.1088\,d$^{-1}$ and 0.9019\,d$^{-1}$ in this F0 star with $T_{\rm eff}$ of 6850\,K. 

\begin{figure}
\centering
\includegraphics[width=0.8\textwidth]{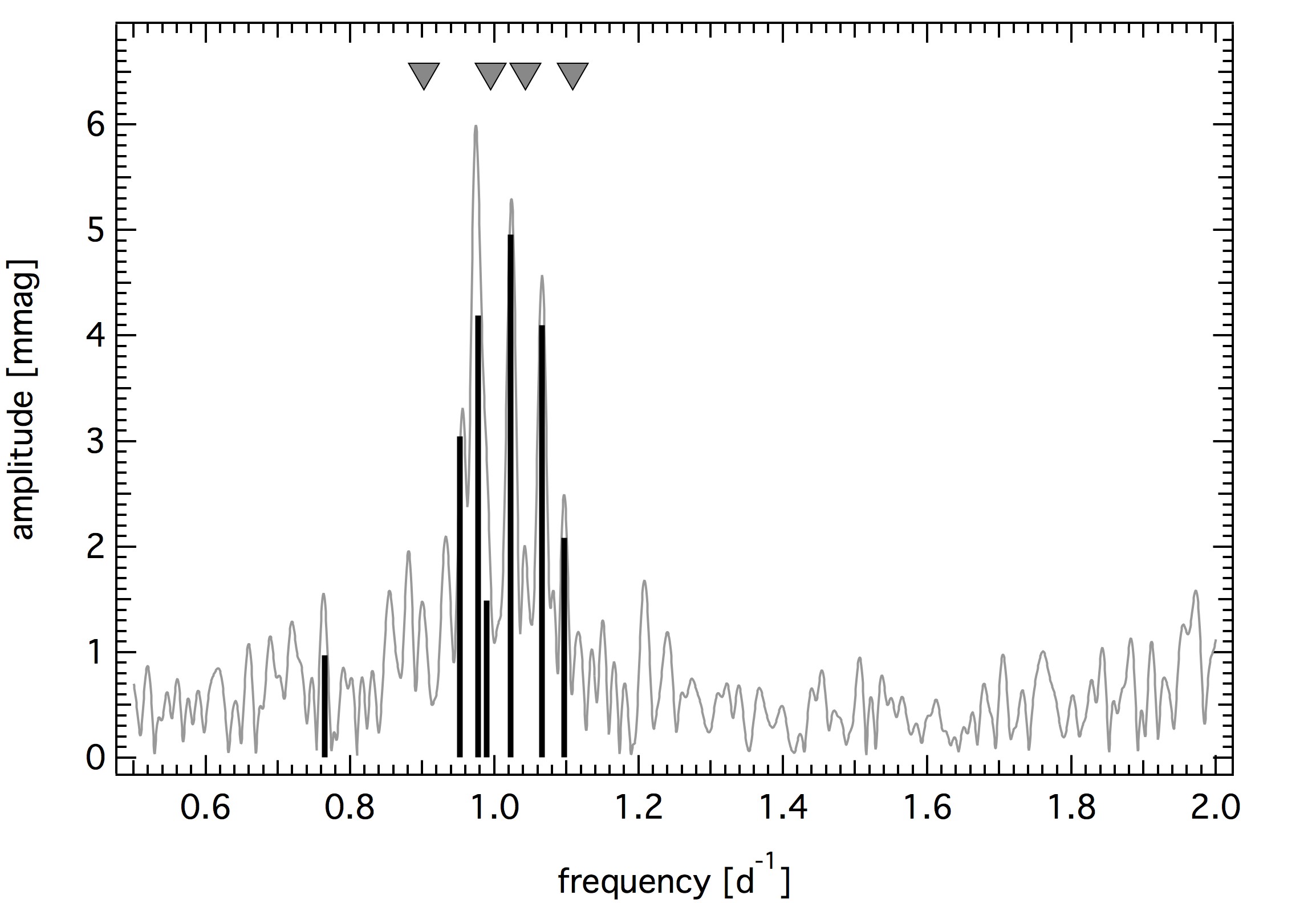}
\caption{Amplitude spectrum of the Heweliusz data for QW Pup (grey), where the seven pulsational frequencies detected in the BRITE-Constellation data are marked in black and the triangles show the frequencies found by \citet{poretti97}.}
\label{qwpup}
\end{figure}

The star was included in the $\beta$ Pictoris - I field in 2015 and observed by Heweliusz for $\sim$ 78 days. From these data we identified seven pulsation frequencies in the range between 0.7 and 1.2\,d$^{-1}$ (see Figure \ref{qwpup}) basically confirming the previous findings. Additional BRITE-Constellation observations of QW Puppis are planned for the period November 2016 to April 2017. A detailed asteroseismic investigation and a search for $g$-mode period spacing patterns will be conducted with the future, significantly longer data sets in both colors.

\subsection{$\beta$ Pictoris (HR 2020, HD 39060)}

The young and nearby (distance $\sim$20\,pc) A-type star $\beta$ Pictoris has been studied frequently in the past and at multiple wavelengths. The star hosts a wide, edge-on, dense circumstellar disk that has been subject of many studies for the last three decades \citep[for an overview see e.g.,][]{lagrange10}, because its structure was theoretically linked to the presence of one or more massive planets. As a final confirmation of the theoretical predictions, \citet{lagrange10} discovered a giant planet ($\beta$ Pictoris b) through direct imaging using NaCo at the ESO VLT.

The age of the $\beta$ Pictoris system has been inferred using various methods \citep[e.g.,][]{mamajek14} and is estimated to lie between 11.5 and 26 million years. The full formation of a planetary system from a disk around a young star needs about 10 million years \citep{mamajek14}. As such, the $\beta$ Pictoris system can be used as a clock to test planet formation processes, but the lower limit of $\beta$ Pictoris' age range appears to be on the lower side to have a fully developed planet. Given the system's youth and close distance, therefore its importance in exoplanetology, the rather large age spread is unsatisfactory and has a profound impact on our understanding of planet formation and early evolution. 

$\beta$ Pictoris is known to be a $\delta$ Scuti pulsator (Koen et al. 2003a, MNRAS 341, 1385). Therefore, asteroseismology can be used to better characterize the star and provide an independent estimate of its age. 
Pulsations in $\beta$ Pictoris were first discovered using photometric measurements obtained with the SAAO 0.5m and 0.75m telescopes \citep{koen03ph}. Using data collected in only four observing nights, three pulsation frequencies were discovered at 47.055\,d$^{-1}$, 38.081 \,d$^{-1}$ and 52.724\,d$^{-1}$ with amplitudes of 1.63, 1.50 and 1.07 mmag, respectively. On the basis of this discovery, \citet{koen03sp} obtained 697 high-dispersion spectra with GIRAFFE at the SAAO 1.7m telescope over a period of two weeks and found 18 pulsation frequencies from the line profile variations. \citet{koen03sp} tried to identify $\ell$ and $m$ for $\beta$ Pictoris reporting the detection of two modes of low degree $\ell$ and 16 other modes that they could not identify. At that time the theoretical models for $\delta$ Scuti pulsations did not distinguish between the pre- and post-MS evolutionary stages. The discovery of pulsations in such a presumably young object was important by itself, and the pulsation frequencies were interpreted as coming from a more evolved post-MS star.

A first try run with BRITE-Constellation on $\beta$ Pictoris was conducted between March and May 2015 using the satellite Heweliusz operating with a red filter. The reduced BRITE-Constellation data from that first try run revealed already several additional pulsation modes that were previously unknown and confirmed some of the findings from 2003. The upcoming BRITE-Constellation campaign on $\beta$ Pictoris between November 2016 and April 2017 will deliver photometric time series in two colors measured by at least two of the five nano-satellites simultaneously.

\section{Conclusions}
BRITE-Constellation observations of $\delta$ Scuti and $\gamma$ Doradus stars have been quite successful in the past. As $\delta$ Scuti pulsations are easier to measure from ground due to their shorter periods, new space data from BRITE-Constellation tend to confirm previous findings and sometimes increase the numbers of detected frequencies. 

For $\gamma$ Doradus type pulsators, BRITE-Constellation can contribute significantly to the advances in that research field due to the long observational time bases of up to 180 days. Currently, only space photometry obtained by the Kepler satellite \citep{koch10} allows to detect period spacing patterns and conduct an asteroseismic mode identification for $\gamma$ Doradus stars. But we found first evidence that we can detect and analyze similar $g$-mode period spacing patterns using BRITE-Constellation data (e.g., for 43 Cygni, see S. G\"ossl's contribution to these proceedings). 

For both groups of pulsators, the color information from the BRITE-Constellation nano-satellites is a new addition that still has great potential in case of simultaneous observations of a given target in both filters with sufficient precision.

BRITE-Constellation will also be able to expand our knowledge about the young star $\beta$ Pictoris, its pulsations, evolutionary stage and planetary system in particular with the upcoming observing run in 2016/2017 which will cover the time during the transit of the planet's Hill sphere (see also M. Kenworthy's contribution to these proceedings).

\acknowledgements{
Based on data collected by the BRITE-Constellation satellite mission, built, launched and operated thanks to support from the Austrian Aeronautics and Space Agency and the University of Vienna, the Canadian Space Agency (CSA) and the Foundation for Polish Science \& Technology (FNiTP MNiSW) and National Centre for Science (NCN).
K.Z. acknowledges support by the Austrian Fonds zur F\"orderung der wissenschaftlichen Forschung (FWF, project V431-NBL).
}

\bibliographystyle{ptapap}
\bibliography{KZwintz}

\begin{thebibliography}{30}
\providecommand{\natexlab}[1]{#1}
\providecommand{\url}[1]{\texttt{#1}}
\providecommand{\urlprefix}{URL }
\providecommand{\eprint}[2][]{\url{#2}}

\bibitem[{{Aerts} et~al.(2010){Aerts}, {Christensen-Dalsgaard}, \&
  {Kurtz}}]{aerts10}
{Aerts}, C., {Christensen-Dalsgaard}, J., {Kurtz}, D.~W., {Asteroseismology}
  (2010)

\bibitem[{{Bowman} et~al.(2016)}]{bowman16}
{Bowman}, D.~M., et~al., \emph{{Amplitude modulation in {$\delta$} Sct stars:
  statistics from an ensemble study of Kepler targets}}, \emph{\mnras}
  \textbf{460}, 1970 (2016), \eprint{1605.03955}

\bibitem[{{Breger} et~al.(2009){Breger}, {Lenz}, \& {Pamyatnykh}}]{breger09}
{Breger}, M., {Lenz}, P., {Pamyatnykh}, A.~A., \emph{{Towards mode selection in
  {$\delta$} Scuti stars: regularities in observed and theoretical frequency
  spectra}}, \emph{\mnras} \textbf{396}, 291 (2009), \eprint{0812.0856}

\bibitem[{{Breger} et~al.(1993)}]{breger93}
{Breger}, M., et~al., \emph{{Nonradial Pulsation of the Delta-Scuti Star
  Bu-Cancri in the Praesepe Cluster}}, \emph{\aap} \textbf{271}, 482 (1993)

\bibitem[{{Bruntt} et~al.(2007)}]{bruntt07}
{Bruntt}, H., et~al., \emph{{Asteroseismology with the WIRE satellite. I.
  Combining ground- and space-based photometry of the {$\delta$} Scuti star
  {$\epsilon$} Cephei}}, \emph{\aap} \textbf{461}, 619 (2007),
  \eprint{astro-ph/0610539}

\bibitem[{{Che} et~al.(2011)}]{che11}
{Che}, X., et~al., \emph{{Colder and Hotter: Interferometric Imaging of
  {$\beta$} Cassiopeiae and {$\alpha$} Leonis}}, \emph{\apj} \textbf{732}, 68
  (2011), \eprint{1105.0740}

\bibitem[{{Escorza} et~al.(2016)}]{escorza16}
{Escorza}, A., et~al., \emph{{HD 41641: A classical {$\delta$} Sct-type
  pulsator with chemical signatures of an Ap star}}, \emph{\aap} \textbf{588},
  A71 (2016), \eprint{1602.04682}

\bibitem[{{Guzik} et~al.(2000)}]{guzik00}
{Guzik}, J.~A., et~al., \emph{{Driving the Gravity-Mode Pulsations in
  {$\gamma$} Doradus Variables}}, \emph{\apjl} \textbf{542}, L57 (2000)

\bibitem[{{Kaye} et~al.(1999)}]{kaye99}
{Kaye}, A.~B., et~al., \emph{{Gamma Doradus Stars: Defining a New Class of
  Pulsating Variables}}, \emph{\pasp} \textbf{111}, 840 (1999),
  \eprint{astro-ph/9905042}

\bibitem[{{Koch} et~al.(2010)}]{koch10}
{Koch}, D.~G., et~al., \emph{{Kepler Mission Design, Realized Photometric
  Performance, and Early Science}}, \emph{\apjl} \textbf{713}, L79-L86 (2010),
  \eprint{1001.0268}

\bibitem[{{Koen}(2003)}]{koen03ph}
{Koen}, C., \emph{{{$\delta$} Scuti pulsations in {$\beta$} Pictoris}},
  \emph{\mnras} \textbf{341}, 1385 (2003)

\bibitem[{{Koen} et~al.(2003)}]{koen03sp}
{Koen}, C., et~al., \emph{{Pulsations in {$\beta$} Pictoris}}, \emph{\mnras}
  \textbf{344}, 1250 (2003)

\bibitem[{{Kurtz} et~al.(2014)}]{kurtz14}
{Kurtz}, D.~W., et~al., \emph{{Asteroseismic measurement of surface-to-core
  rotation in a main-sequence A star, KIC 11145123}}, \emph{\mnras}
  \textbf{444}, 102 (2014), \eprint{1405.0155}

\bibitem[{{Kuschnig} et~al.(1997)}]{kuschnig97}
{Kuschnig}, R., et~al., \emph{{Microvariability survey with the Hubble Space
  Telescope Fine Guidance Sensors. Exploring the instrumental properties}},
  \emph{\aap} \textbf{328}, 544 (1997)

\bibitem[{{Lagrange} et~al.(2010)}]{lagrange10}
{Lagrange}, A.-M., et~al., \emph{{A Giant Planet Imaged in the Disk of the
  Young Star {$\beta$} Pictoris}}, \emph{Science} \textbf{329}, 57 (2010),
  \eprint{1006.3314}

\bibitem[{{Lenz} \& {Breger}(2005)}]{period04}
{Lenz}, P., {Breger}, M., \emph{{Period04 User Guide}}, \emph{Communications in
  Asteroseismology} \textbf{146}, 53 (2005)

\bibitem[{{Mamajek} \& {Bell}(2014)}]{mamajek14}
{Mamajek}, E.~E., {Bell}, C.~P.~M., \emph{{On the age of the {$\beta$} Pictoris
  moving group}}, \emph{\mnras} \textbf{445}, 2169 (2014), \eprint{1409.2737}

\bibitem[{{Miglio} et~al.(2008){Miglio}, {Montalb{\'a}n}, {Noels}, \&
  {Eggenberger}}]{miglio08}
{Miglio}, A., {Montalb{\'a}n}, J., {Noels}, A., {Eggenberger}, P.,
  \emph{{Probing the properties of convective cores through g modes: high-order
  g modes in SPB and {$\gamma$} Doradus stars}}, \emph{\mnras} \textbf{386},
  1487 (2008), \eprint{0802.2057}

\bibitem[{{Neiner} \& {Lampens}(2015)}]{neiner15}
{Neiner}, C., {Lampens}, P., \emph{{First discovery of a magnetic field in a
  main-sequence {$\delta$} Scuti star: the Kepler star HD 188774}},
  \emph{\mnras} \textbf{454}, L86 (2015), \eprint{1508.07273}

\bibitem[{{Poretti} et~al.(1997)}]{poretti97}
{Poretti}, E., et~al., \emph{{Discovery and analysis of Gamma Doradus type
  pulsations in the F0 IV star HR 2740=QW PUP}}, \emph{\mnras} \textbf{292},
  621 (1997)

\bibitem[{{Reegen}(2007)}]{reegen07}
{Reegen}, P., \emph{{SigSpec. I. Frequency- and phase-resolved significance in
  Fourier space}}, \emph{\aap} \textbf{467}, 1353 (2007),
  \eprint{physics/0703160}

\bibitem[{{Reese} et~al.(2009)}]{reese09}
{Reese}, D.~R., et~al., \emph{{Pulsation modes in rapidly rotating stellar
  models based on the self-consistent field method}}, \emph{\aap} \textbf{506},
  189 (2009), \eprint{0903.4854}

\bibitem[{{Riboni} et~al.(1994){Riboni}, {Poretti}, \& {Galli}}]{riboni94}
{Riboni}, E., {Poretti}, E., {Galli}, G., \emph{{The long-term behaviour of the
  photometric variability of {$\beta$} Cassiopeiae.}}, \emph{\aaps}
  \textbf{108} (1994)

\bibitem[{{Tassoul}(1980)}]{tassoul80}
{Tassoul}, M., \emph{{Asymptotic approximations for stellar nonradial
  pulsations}}, \emph{\apjs} \textbf{43}, 469 (1980)

\bibitem[{{Van Reeth} et~al.(2016){Van Reeth}, {Tkachenko}, \&
  {Aerts}}]{vanreeth16}
{Van Reeth}, T., {Tkachenko}, A., {Aerts}, C., \emph{{The interior rotation of
  a sample of gamma Doradus stars from ensemble modelling of their gravity mode
  period spacings}}, \emph{ArXiv e-prints}  (2016), \eprint{1607.00820}

\bibitem[{{Van Reeth} et~al.(2015{\natexlab{a}})}]{vanreeth15aa}
{Van Reeth}, T., et~al., \emph{{Detecting non-uniform period spacings in the
  Kepler photometry of {$\gamma$} Doradus stars: methodology and case
  studies}}, \emph{\aap} \textbf{574}, A17 (2015{\natexlab{a}}),
  \eprint{1410.8178}

\bibitem[{{Van Reeth} et~al.(2015{\natexlab{b}})}]{vanreeth15apjs}
{Van Reeth}, T., et~al., \emph{{Gravity-mode Period Spacings as a Seismic
  Diagnostic for a Sample of {$\gamma$} Doradus Stars from Kepler Space
  Photometry and High-resolution Ground-based Spectroscopy}}, \emph{\apjs}
  \textbf{218}, 27 (2015{\natexlab{b}}), \eprint{1504.02119}

\bibitem[{{Zwintz} et~al.(2011)}]{zwintz11}
{Zwintz}, K., et~al., \emph{{Regular frequency patterns in the classical
  {$\delta$} Scuti star HD 144277 observed by the MOST satellite}}, \emph{\aap}
  \textbf{533}, A133 (2011), \eprint{1109.2743}

\bibitem[{{Zwintz} et~al.(2013)}]{zwintz13}
{Zwintz}, K., et~al., \emph{{Regular frequency patterns in the young {$\delta$}
  Scuti star HD 261711 observed by the CoRoT and MOST satellites}}, \emph{\aap}
  \textbf{552}, A68 (2013), \eprint{1302.3369}

\bibitem[{{Zwintz} et~al.(2014)}]{zwintz14}
{Zwintz}, K., et~al., \emph{{Echography of young stars reveals their
  evolution}}, \emph{Science} \textbf{345}, 550 (2014), \eprint{1407.4928}

\end{thebibliography}

\end{document}